# Derivation of the coefficient squared probability law in quantum mechanics.

Casey Blood
Professor Emeritus of Physics, Rutgers University
CaseyBlood@gmail.com

**Abstract.**
If one assumes there is probability of perception in quantum mechanics, then unitarity dictates that it must have the coefficient squared form, in agreement with experiment.

There are two parts to the highly successful theory of quantum mechanics. First, there are the Schrödinger equations of motion which give accurate descriptions of a wide range of phenomena. Second, there is the coefficient squared probability law which says that the probability of perceiving an outcome corresponding to state $|j\rangle$ is equal to $|a(j)|^2$. This is a separate law, not derivable from the Schrödinger equations (see below). In fact, there is no probability at all inherent in the Schrödinger equations of motion. Here we make certain assumptions about probability and then give a derivation of the coefficient squared ( $|a(j)|^2$ ) form of the law. Zurek [1-3] derives the same result using different assumptions and a different, somewhat more complicated method.

**1. Basic quantum mechanics and its successes.**
We start with a summary of *basic* quantum mechanics and what it can account for. In basic quantum mechanics (1) only the state vectors exist (no particles, no observers other than the quantum mechanical versions of the observers), (2) there are no hidden variables, (3) there is no collapse, and (4) the time translation operator is linear and unitary. This is essentially Everett's many-worlds [4] scheme, but without his treatment of probability.

Basic quantum mechanics is highly successful. In addition to correctly predicting many of the quantitative properties of matter—energy levels and so on—it also accounts for the more qualitative aspects of our perceptions.

- With no collapse or particles, it accounts for the perception of just one, 'classical' outcome to a measurement (the measurement 'problem') [5].
- It accounts for the perception of only one exposed grain of film for each photon-like wave function that goes through a double slit [6].
- It accounts for particle-like trajectories [7].
- Through group representation theory, it implies that the particle-like properties of matter—mass, energy, momentum, spin, and charge—are all properties of the state vectors [6].
- It accounts for the photoelectric effect and Compton scattering [6].

That is, it accounts for all the particle-*like* properties of matter [6]. Thus, because the wave functions/state vectors can account for all the particle-like properties of matter, there is no evidence for particles; wave-particle duality is just a duality in the properties—wave-*like* and particle-*like*—of the wave functions.

•In addition, with no need to invoke action at a distance, it accounts for all the entangled state experiments—the Bell-Aspect experiment [8,9], the quantum eraser [10], the Ionicioiu-Terno version of the Wheeler delayed-choice experiment [11], and so on.

## 2. Basic quantum mechanics cannot account for probability.

Probability, however, is the downfall of basic quantum mechanics as a complete interpretation because, by itself, the theory contains no probability at all. To be more specific, suppose there are N different possible outcomes of an experiment. Then after a single run, there are N isolated versions of reality. All versions are equally valid in the sense that no version of the outcome is singled out in the mathematics of basic quantum mechanics as the only 'real' one, the one corresponding to what we perceive. And there is no singular "I" in basic quantum mechanics whose perceptions correspond to just one version of the observer. Thus, because there is no singling out of any sort, there can be no probability of perceiving a given version of reality. (Note: We use *perception* of a given outcome rather than *existence* of a given outcome because we don't know that only one outcome 'actually exists.')

It is not our purpose here to consider how basic quantum mechanics is modified to accommodate probability of perception. We will simply assume that, for some currently not-understood reason, when the wave function contains several potential versions of reality, there is a probability of perceiving a given outcome (Zurek [1-3] also makes this assumption). We will then show that the $|a(j)|^2$ form of the law can be derived by using the unitarity of basic quantum mechanics plus two additional, relatively weak assumptions.

## 3. Derivation of the $|a(j)|^2$ probability law for a three-state system.

We start with a three-state system

$$|\psi\rangle = a(1)|1\rangle + a(2)|2\rangle + a(3)|3\rangle \tag{1}$$

$$|a(1)|^2 + |a(2)|^2 + |a(3)|^2 = 1 \tag{2}$$

where Eq. (2) comes from **unitarity**. The aim is to determine an equation for the probability of perceiving each state. We assume the experiment is akin to a spin 1 Stern-Gerlach experiment and that there is a detector on each of the three paths. The phase factors on the *a*'s can be absorbed into the definition of the three detector wave functions, so we can assume the *a*'s are real. We then suppose the probability of observing outcome *j* is

$$Prob_\psi(|j\rangle) = P_j(a(j), v) \tag{3}$$

That is, **$P_j$ is presumed to be a function of $a(j)$** (but none of the other *a*'s), with the possibility that it is a different function for each *j*. The *v* represents anything besides the coefficients that the probability might depend on. (It might depend on the specifics of the detectors used, for example.)

The probabilities obey the following equations.

$$\sum_{j=1}^{3} P_j(a(j), v) = 1 \tag{4}$$

$$P_j(0, v) = 0 \quad (5)$$
$$P_j(1, v) = 1$$

The first assumption is inherent in the idea of probability. The other two say the state will never be observed if it does not occur in $|\psi\rangle$; and it will always be observed if it is the only state occurring in $|\psi\rangle$.

We now assume $P_j(a(j), v)$ is a **differentiable function** of $a(j)$ and take the variation of Eq. (4) with respect to the $a(j)$s, subject to the constraint of Eq. (2). We use the calculus of variations, with Lagrange multiplier $\lambda$, to obtain

$$\frac{\partial P_j}{\partial a(j)} = 2a(j)\lambda \quad (6)$$

If we now use the constraints of Eqs. (4) and (5) to determine the constants of integration and $\lambda$, we get

$$P_j(a(j), v) = |a(j)|^2 \quad (7)$$

That is, we get just the experimentally confirmed $|a(j)|^2$ rule. Note that the assumptions imply that probability depends only on the coefficients; it is independent of any other parameters (the $v$). Basically, once one assumes there is a probability, the condition of unitarity completely determines the $|a(j)|^2$ rule!

### 4. Two-state system.

The above argument works when the system has three or more possible outcomes. But it doesn't work when there are just two outcomes (because a(2) cannot be varied independently of a(1)) so we need a slightly different argument. We will give it for a partially silvered mirror experiment. Photons are incident on a partially silvered mirror, with one part of the resulting wave function, $|H\rangle$, traveling on a horizontal path and the other part, $|V\rangle$, on a vertical path. The vertical beam is further split into two parts, $|x\rangle$ and $|y\rangle$, so the state vector is

$$|\psi\rangle = a|H\rangle + c|x\rangle + d|y\rangle \quad (8)$$

$$a^2 + c^2 + d^2 = 1 \quad (9)$$

We are now back to a three-state system so the arguments of Sec. 3 apply, with the result again being the coefficient squared rule for probability.

### 5. Summary.

We assume the following:
(1). If we have a state $\sum a(j)|j\rangle$, there is a probability, $P_j$, of perceiving outcome $j$.
(2). The probability $P_j$ is a differentiable function of coefficient $a(j)$. It may also be a function of other variables $v$ (but not the other coefficients).
(3). The time evolution of states is unitary.

It then follows that probability obeys the experimentally confirmed formula, $P_j = |a(j)|^2$. So we see that the primary mystery is not the $|a(j)|^2$ form of the probability law, but the origin in quantum mechanics of probability itself.